\documentclass[a4paper,11pt]{article}
\pdfoutput=1 

\usepackage{jheppub} 

\usepackage[T1]{fontenc} 
\usepackage{xcolor} 
\usepackage{hyperref} 
\usepackage{enumitem}

\usepackage{booktabs}

\usepackage{subcaption}

\usepackage{multirow} 

\usepackage{tabularx} 

\usepackage{makecell}

\title{\boldmath Primordial Gravitational Waves of Big Bounce Cosmology in Light of Stochastic Gravitational Wave Background}

\author{Changhong Li}

\affiliation{Department of Astronomy, Key Laboratory of Astroparticle Physics of Yunnan Province,\\  School of Physics and Astronomy,  Yunnan University,\\ No.2 Cuihu North Road, Kunming,  650091 China}
\emailAdd{changhongli@ynu.edu.cn}

\abstract{Primordial gravitational waves from the very early stages of the universe, such as inflation or bounce processes, are an irreducible cosmological source of the stochastic gravitational wave background (SGWB). The recent detection of SGWB signals around the nano-Hertz frequency by pulsar timing arrays (PTAs), including NANOGrav, EPTA, PPTA, IPTA, and CPTA, opens a new window to explore these very early stages of the universe through these primordial gravitational waves. In this work, we investigate the generation and evolution of primordial gravitational waves in a generic big bounce cosmology by parameterizing its background evolution into four phases, where perturbation modes exit and re-enter the horizon twice. By analytically solving the equation of motion for primordial gravitational waves and matching solutions at the boundaries, we obtain the explicit form of the primordial gravitational wave spectrum in a generic big bounce cosmology. We find that, according to the evolution of primordial gravitational waves, a generic scenario of big bounce cosmology can be categorized into four distinct types. We introduce four toy models for these categories, demonstrating that our analytical results can be straightforwardly applied to various bouncing universe models in which the equation of state of the background is constant in each phase. We also prospect future applications of our results in interpreting SGWB signals searched by PTAs and upcoming advanced gravitational wave detectors such as SKA, Taiji, Tianqin, LISA, DECIGO, and aLIGO/Virgo/KAGRA using Bayesian analysis.}

\begin{document}

\maketitle

\flushbottom

\section{Introduction}    \label{sec:intro}  
The standard slow-roll inflation is renowned for resolving the initial problems of the Big Bang Theory: the horizon problem and flatness problem~\cite{Guth:1980zm, Starobinsky:1980te, Sato:1980yn, Linde:1981mu, Albrecht:1982wi, Mukhanov:1990me}. However, it suffers from its own initial problem: the initial singularity problem, which undermines its validity~\cite{Borde:1993xh, Borde:2001nh}. To address this issue, the bounce universe scenario, in which the universe transitions from contraction to expansion with a non-zero minimal size, has been proposed and extensively studied~\cite{Khoury:2001wf, Gasperini:2002bn, Creminelli:2006xe, Peter:2006hx, Cai:2007qw, Cai:2008qw, Saidov:2010wx, Li:2011nj, Cai:2011tc, Easson:2011zy, Bhattacharya:2013ut, Qiu:2015nha, Cai:2016hea, Barrow:2017yqt, deHaro:2017yll, Ijjas:2018qbo, Boruah:2018pvq, Nojiri:2019yzg} (for comprehensive reviews, see~\cite{Novello:2008ra, Brandenberger:2016vhg, Nojiri:2017ncd, Odintsov:2023weg} and references therein). Akin to inflation, the bounce universe can also generate a nearly scale-invariant primordial curvature perturbation compatible with current observations of cosmic microwave background (CMB) anisotropies~\cite{WMAP:2010qai, Planck:2015fie, Planck:2018vyg}, making it a compelling alternative to the standard inflation scenario. Furthermore, in late-time cosmology, the bounce universe can produce dark matter particles with an energy density fraction compatible with current astrophysical observations, $\Omega_\chi = 0.26$~\cite{Li:2014era, Cheung:2014nxi, Li:2014cba, Li:2015egy, Li:2020nah}. Therefore, it is sometimes recognized as "big bounce cosmology"~\cite{Nojiri:2017ncd}.

Currently, an array of pulsar timing array (PTA) experiments, including NANOGrav \cite{NANOGrav:2020spf, NANOGrav:2023gor}, EPTA \cite{EPTA:2021crs, EPTA:2023fyk}, PPTA \cite{Goncharov:2021oub, Reardon:2023gzh}, IPTA \cite{Antoniadis:2022pcn}, and CPTA \cite{Xu:2023wog}, have detected a stochastic gravitational wave background (SGWB) around the nano-Hertz frequency. Besides other astrophysical and cosmological sources of SGWB, including supermassive black hole binaries, cosmic strings, domain walls, first-order phase transitions, scalar-induced gravitational waves, audible axion, kination-domination, preheating, and particle production~\cite{Sesana:2013wja, Kelley:2017lek, Chen:2018znx, Burke-Spolaor:2018bvk, Siemens:2006yp, Cui:2018rwi, Gouttenoire:2019kij, Hiramatsu:2013qaa, Ferreira:2022zzo,Bian:2022qbh, Gouttenoire:2023ftk, Hindmarsh:2013xza, Hindmarsh:2015qta, Caprini:2015zlo, Hindmarsh:2017gnf, Gouttenoire:2023bqy, Salvio:2023ynn, Ananda:2006af, Baumann:2007zm, Kohri:2018awv, Machado:2019xuc, Machado:2018nqk, Co:2021lkc, Oikonomou:2023qfz, Bethke:2013aba, Adshead:2019igv, Dimastrogiovanni:2016fuu, DAmico:2021fhz}, this finding opens a new window to explore these very early stages of the universe, such as inflation or bounce processes, since primordial gravitational waves from these stages are an irreducible cosmological source of SGWB~\cite{Caprini:2018mtu}. So far, significant efforts have been made to interpret PTA data on SGWB signals within the inflationary framework~\cite{Zhao:2013bba, Guzzetti:2016mkm, Vagnozzi:2020gtf, Vagnozzi:2023lwo} (for recent reviews, see \cite{NANOGrav:2023hvm, EPTA:2023xxk, Figueroa:2023zhu, Bian:2023dnv, Ellis:2023oxs} and references therein). However, the interpretation of PTA data on SGWB signals in the context of the bounce universe~\cite{ Boyle:2004gv, Piao:2004jg, Cai:2016hea} remains rare and sporadic~\cite{Zhu:2023lbf, Papanikolaou:2024fzf,  Li:2024oru, Ben-Dayan:2024aec}, partially due to the complexity of the multiple phases in the cosmic evolution of the bounce universe.

\begin{figure}[htbp]
\centering 
\includegraphics[width=1.0\textwidth]{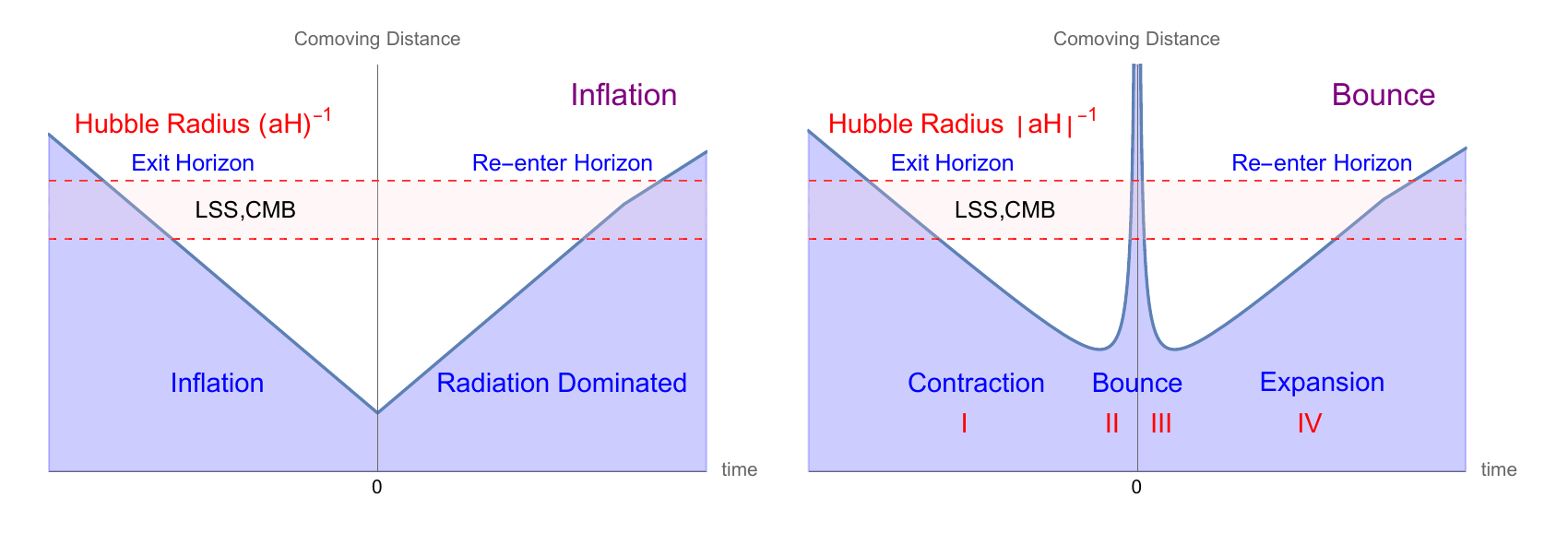}
\caption{\label{fig:Infboure.pdf} Cosmic evolution of the effective Hubble radius in generic inflationary (left) and bouncing (right) scenarios.}
\end{figure}

To pave the way for investigating the bounce universe interpretation of PTA data on SGWB signals, we systematically compute the generation and cosmic evolution of primordial gravitational waves throughout the entire cosmic history of a generic big bounce cosmology. As illustrated in Fig.~\ref{fig:Infboure.pdf}, although the cosmic backgrounds of inflation and the bounce universe are entirely different, the evolution of their effective Hubble radius, $\left|aH\right|^{-1}$, is strikingly similar~\cite{Cheung:2016vze}. This implies that they have a very similar and dual pattern in generating primordial curvature perturbations~\cite{Wands:1998yp, Finelli:2001sr, Boyle:2004gv, Raveendran:2023auh}. However, it should be noted that, in contrast to the simple pattern in inflation, the effective Hubble radius has a spike around the bounce point in big bounce cosmology. This indicates that, to explicitly compute the spectrum of primordial gravitational waves in the bounce universe, we must take into account all four phases: collapsing contraction (Phase I), bouncing contraction (Phase II), bouncing expansion (Phase III), and decelerating expansion (Phase IV), rather than only two phases in inflationary cosmology.

Assuming the equation of state (EoS) of the cosmic background is constant in each phase, we solve the equation of motion (EoM) for tensor modes of metric perturbation (primordial gravitational waves) and match their solutions at the boundaries of each phase. Eventually, we obtain an analytical expression for the spectrum of primordial gravitational waves in a generic big bounce cosmology. We find that, according to the evolution of primordial gravitational waves, a generic scenario of big bounce cosmology can be categorized into four types. We introduce four toy models for these categories, demonstrating that our analytical results can be straightforwardly applied to various bouncing universe models, which can be used to interpret SGWB signals detected by PTAs in further studies.

This paper is organized as follows: In Section 2, we parameterize the cosmic evolution of a generic big bounce cosmology. In Section 3, we solve the EoM for primordial gravitational waves and match them at their boundaries to obtain the explicit form of the spectrum. In Section 4, we categorize big bounce cosmologies based on the evolution of primordial gravitational waves and introduce corresponding models to demonstrate these categorizations. In Section 5, we summarize our results and discuss potential future applications, particularly concerning the SGWB signals detected by PTAs.

\section{Cosmic Evolution of a Generic Big Bounce Cosmology}

The cosmic evolution of a generic big bounce cosmology consists of four phases: collapsing contraction (Phase I), bouncing contraction (Phase II), bouncing expansion (Phase III), and decelerating expansion (Phase IV). As illustrated in Fig.~\ref{fig:Infboure.pdf}, the evolution of the effective Hubble radius is different in these four phases. Using $a(\eta)=a_i\eta^{\nu_i}$, we parameterize these four phases as:

\begin{enumerate}[label=\Roman*.]
    \item {\bf Collapsing contraction phase} ($\dot{a}<0$ and $\ddot{a}<0$). During this phase, the universe contracts with increasing speed ($\ddot{a}/\dot{a}>0$), which corresponds to
    \begin{equation}
        \nu_{1}>0~, \quad \eta:\infty\rightarrow 0~,
    \end{equation}
    where $a_i$ and $\nu_i$ are the coefficient and power-law index of the scale factor $a(\eta)$ in terms of conformal time $\eta$, respectively. We use the subscript $i=1,2,3,4$ to denote each phase. During this phase, the effective Hubble radius decreases, so the primordial gravitational modes exit the horizon ($k\eta\simeq 1$) from sub-horizon ($k\eta\gg 1$) to super-horizon ($k\eta\ll 1$), as shown in I in Fig.~\ref{fig:Infboure.pdf}.

    \item {\bf Bouncing contraction phase} ($\dot{a}<0$ and $\ddot{a}>0$). In this phase, the speed of cosmic contraction decreases ($\ddot{a}/\dot{a}<0$), which corresponds to
    \begin{equation}
        \nu_{2}<0~, \quad \eta:0\rightarrow \infty~.
    \end{equation}
    In this phase, the effective Hubble radius increases, so the primordial gravitational modes re-enter the horizon ($k\eta\simeq 1$) from super-horizon ($k\eta\ll 1$) to sub-horizon ($k\eta\gg 1$), as shown in II in Fig.~\ref{fig:Infboure.pdf}.

    \item {\bf Bouncing expansion phase} ($\dot{a}>0$ and $\ddot{a}>0$). In this phase, the universe expands from the bounce point with accelerating speed ($\ddot{a}/\dot{a}>0$), which corresponds to 
    \begin{equation}
        \nu_{3}<0, \quad \eta:\infty\rightarrow 0.
    \end{equation}
    During this phase, the effective Hubble radius decreases again, so the primordial gravitational modes exit the horizon ($k\eta\simeq 1$) again from sub-horizon ($k\eta\gg 1$) to super-horizon ($k\eta\ll 1$), as shown in III in Fig.~\ref{fig:Infboure.pdf}.

    \item {\bf Decelerating expansion phase} ($\dot{a}>0$ and $\ddot{a}<0$). In this phase, the universe evolves into a decelerating expansion. For example, the post-reheating era of the standard model of cosmology ($\Lambda$CDM model) is dominated by normal radiation and matter, akin to the conventional inflation model. In this phase, the cosmic expansion decelerates ($\ddot{a}/\dot{a}<0$), which corresponds to
    \begin{equation}
        \nu_{4}>0~, \quad \eta:0\rightarrow \infty~.
    \end{equation}
    In this phase, the effective Hubble radius increases again, so the primordial gravitational modes re-enter the horizon ($k\eta\simeq 1$) again from super-horizon ($k\eta\ll 1$) to sub-horizon ($k\eta\gg 1$), as shown in IV in Fig.~\ref{fig:Infboure.pdf}.

\end{enumerate}

In Table~\ref{tab:fourphpar}, we summarize the parameter region of each phase in big bounce cosmology, where $k$ is the comoving wavevector of primordial gravitational modes and $\tilde{\nu} \equiv |\nu - 1/2|$ is a parameter introduced to facilitate computation.

\begin{table}[ht]
    \centering
    \caption{Four phases in big bounce cosmology and their parameter regions} \label{tab:fourphpar}
    \begin{tabular}{lllllll}
    \toprule[0.4mm]
        \makecell[c]{Phase} &$ \makecell[c]{~\dot{a}} $& $\makecell[c]{~\ddot{a}}$ & $\makecell[c]{~w}$ & $\makecell[c]{~\nu}$ & $\makecell[c]{~k\eta}$ & \makecell[c]{$\tilde{\nu}\equiv |\nu-1/2|$} \\ \hline
         Collapsing contraction & ~$<0$ & ~$<0$ & ~$>-1/3$ & ~$>0$ & ~$\to 0$ & ~\makecell[c]{$\tilde{\nu}> 0$} \\
        Bouncing contraction & ~$<0$ & ~$>0$ & ~$<-1/3$ & ~$ < 0$ & ~$\to \infty$ & ~\makecell[c]{$\tilde{\nu}< \frac{1}{2}$} \\
         Bouncing expansion & ~$>0$ & ~$>0$ & ~$<-1/3$ & ~$< 0$ & ~$\to 0$ & ~\makecell[c]{$\tilde{\nu}< \frac{1}{2}$} \\
          Decelerating expansion & ~$>0$ & ~$<0$ & ~$>-1/3$ & ~$>0$ & ~$\to \infty$ & ~\makecell[c]{$\tilde{\nu}> 0$}  
        \\ \toprule[0.4mm]
    \end{tabular}
\end{table}

\section{Primordial Gravitational Waves in Big Bounce Cosmology}

With the parameterization of cosmic evolution in big bounce cosmology, we can solve the EoM of primordial gravitational waves in each phase and match their solutions at the boundaries to obtain the final spectrum.

\subsection{Equation of Motion and General Solution}
The perturbed FLRW metric for studying primordial gravitational waves takes the form:
\begin{equation}
    ds^2 = a^2(\eta)\left[-d\eta^2 + (\delta_{ij} + h_{ij}) dx^i dx^j\right]~,
\end{equation}
where $h_{ij}$ is the transverse and traceless tensor perturbation, which can be decomposed into two polarization states, $h_+$ and $h_\times$. Following~\cite{Caprini:2018mtu}, we denote these two modes as $h_r$ with $r=+,\times$.

Expanding the Einstein equations to first order, we obtain the equation of motion for the tensor mode of metric perturbation:
\begin{equation}\label{eq:eomvk}
    u_k^{\prime\prime} + \left(k^2 - \frac{a^{\prime\prime}}{a}\right) u_k = 0,
\end{equation}
where $u_k$ is the Fourier mode with wave-vector $k$ of the normalized primordial gravitational wave, $u_r \equiv \frac{1}{\sqrt{16\pi G}} a(\eta) h_r$, and $^\prime$ denotes the derivative with respect to $\eta$.

Expressing the scale factor $a(\eta)$ in terms of conformal time,
\begin{equation}
a(\eta)=a_i \eta^{\nu_i},    
\end{equation} 
we obtain the equation of motion for the normalized primordial gravitational wave,
\begin{equation}\label{eq:eomvksim}
    u_k^{[i]\prime\prime} + \left[k^2 - \frac{1}{\eta^2}\left(\tilde{\nu}_i^2 - \frac{1}{4}\right)\right] u_k^{[i]} = 0~,
\end{equation}
where the superscript $^{[i]}$ labels each phase with $i=1,2,3,4$ respectively, and 
\begin{equation}
    \tilde{\nu}_i \equiv \left|\nu_i - \frac{1}{2}\right|~.
\end{equation}

The general solution of the normalized mode of the primordial gravitational wave, Eq.~\ref{eq:eomvksim}, is given by
\begin{align}\label{eqsi:Hankelsols}
    u_k^{[i]} = \sqrt{\eta}\left[A_i H_{\tilde{\nu}_i}^{(1)}(k\eta) + B_i H_{\tilde{\nu}_i}^{(2)}(k\eta)\right]~,
\end{align}
where $H_{\tilde{\nu}_i}^{(1)}(k\eta)$ and $H_{\tilde{\nu}_i}^{(2)}(k\eta)$ are Hankel functions of the first and second kind, respectively.

\subsection{Solutions in sub-horizon and super-horizon limits}

To match the solutions at the boundaries of phases, we need to use the asymptotic solution in either the sub-horizon or super-horizon limits.

{\bf In the sub-horizon limit ($k\eta\gg 1$)}, the asymptotic solution of $u_{k}^{[i]}$ takes the form
\begin{align}\label{eqsi:gensolsub}
    u_{k\uparrow}^{[i]} = \sqrt{\frac{2}{\pi k}} \left[ A_i e^{i \left( k \eta - \frac{\tilde{\nu}_i \pi}{2} - \frac{\pi}{4} \right)} + B_i e^{-i \left( k \eta - \frac{\tilde{\nu}_i \pi}{2} - \frac{\pi}{4} \right)} \right], \quad k\eta\gg 1~,
\end{align}
where the subscript $_{\uparrow}$ labels solutions in the $k\eta\gg 1$ limit.

Accordingly, the gravitational wave mode and its conformal velocity, $h_{k\uparrow}^{[i]}$ and $h_{k\uparrow}^{[i]\prime}$, respectively, take the form
\begin{align}
h_{k\uparrow}^{[i]} &= \frac{\sqrt{16\pi G}}{a(\eta)}\sqrt{\frac{2}{\pi k}} \left[ A_i e^{i \left( k \eta - \frac{\tilde{\nu}_i \pi}{2} - \frac{\pi}{4} \right)} + B_i e^{-i \left( k \eta - \frac{\tilde{\nu}_i \pi}{2} - \frac{\pi}{4} \right)} \right], \\
h_{k\uparrow}^{[i]\prime} &= \frac{\sqrt{16\pi G}}{a(\eta)}\sqrt{\frac{2}{\pi k}} \left[ A_i \left(-\frac{\tilde{\nu}_i}{\eta} + ik \right) e^{i \left( k \eta - \frac{\tilde{\nu}_i \pi}{2} - \frac{\pi}{4} \right)} + B_i \left(-\frac{\tilde{\nu}_i}{\eta} - ik \right) e^{-i \left( k \eta - \frac{\tilde{\nu}_i \pi}{2} - \frac{\pi}{4} \right)} \right],
\end{align}
where $h_k^{[i]}(\eta) = \sqrt{16\pi G} \, a^{-1}(\eta) \, u_k^{[i]}(\eta)$.

{\bf In the super-horizon limit ($k\eta\ll 1$)}, the asymptotic solution of $u_k^{[i]}$ takes the form
\begin{align}\label{eqsi:gensolsup}
    u_{k\downarrow}^{[i]} = \sqrt{\eta}\left[E_i (k\eta)^{\tilde{\nu}_i} + F_i (k\eta)^{-\tilde{\nu}_i}\right], \quad k\eta \ll 1~,
\end{align}
where the subscript $_{\downarrow}$ labels solutions in the $k\eta \ll 1$ limit. The coefficients $(E_i, F_i)$ are given by
\begin{equation}\label{eqsi:tranmatr}
    \begin{pmatrix}
        E_i\\
        F_i
    \end{pmatrix} = T_i
    \begin{pmatrix}
        A_i\\
        B_i
    \end{pmatrix}~,
\end{equation}
with a transformation matrix $T_i$,
\begin{equation}
    T_i \equiv 
    \begin{pmatrix}
        \alpha_i & \alpha_i^\ast \\
        \beta_i & \beta_i^\ast
    \end{pmatrix}~.
\end{equation}
The elements $\alpha_i$ and $\beta_i$ are given by
\begin{equation}\label{eqsi:alphaexp}
    \alpha_i \equiv 2^{-\tilde{\nu}_i} \left[\frac{1}{\Gamma(\tilde{\nu}_i + 1)} - i\frac{\Gamma(-\tilde{\nu}_i) \cos(\pi \tilde{\nu}_i)}{\pi}\right] = 2^{-\tilde{\nu}_i} \left[\frac{-ie^{i\pi \tilde{\nu}_i}}{\sin(\pi \tilde{\nu}_i)\Gamma(\tilde{\nu}_i + 1)}\right]~,
\end{equation}
\begin{equation}
    \beta_i \equiv -i \left[ \frac{\Gamma(\tilde{\nu}_i)}{\pi} 2^{\tilde{\nu}_i} \right]~.
\end{equation}
Here, $\Gamma(z) = \int_0^\infty t^{z-1} e^{-t} dt$ is the Gamma function. To derive the second expression in Eq.~\ref{eqsi:alphaexp}, we used the reflection formula for the Gamma function: $\Gamma(z)\Gamma(1-z) = \pi/\sin{(\pi z)}$.

Accordingly, the gravitational wave mode and its conformal velocity, $h_{k\downarrow}^{[i]}$ and $h_{k\downarrow}^{[i]\prime}$, respectively, take the form
\begin{align}
    h_{k\downarrow}^{[i]} &= \frac{\sqrt{16\pi G}}{a(\eta)} \sqrt{\eta} \left[E_i (k\eta)^{\tilde{\nu}_i} + F_i (k\eta)^{-\tilde{\nu}_i}\right], \\
    h_{k\downarrow}^{[i]\prime} &= \frac{\sqrt{16\pi G}}{a(\eta) \sqrt{\eta}} \left[E_i \chi_i (k\eta)^{\tilde{\nu}_i} + F_i \bar{\chi}_i (k\eta)^{-\tilde{\nu}_i}\right],
\end{align}
where 
\begin{align}\label{eqsi:chiidef}
    \chi_i &\equiv \tilde{\nu}_i - \left(\nu_i - \frac{1}{2}\right) = \left|\nu_i - \frac{1}{2}\right| - \left(\nu_i - \frac{1}{2}\right), \\
    \bar{\chi}_i &\equiv -\tilde{\nu}_i - \left(\nu_i - \frac{1}{2}\right) = -\left|\nu_i - \frac{1}{2}\right| - \left(\nu_i - \frac{1}{2}\right)~.\label{eqsi:bchiidef}
\end{align}

\subsection{Matching Conditions and Their Matrix Representations}

The matching conditions at the boundary between phases $i$ and $i+1$ are given by:
\begin{equation}\label{eqsi:matconone}
    h_k^{[i+1]}(\eta_{m_i}) = h_k^{[i]}(\eta_{m_i}),
\end{equation}
and 
\begin{equation}\label{eqsi:matcontwo}
    h_k^{\prime [i+1]}(\eta_{m_i}) = h_k^{\prime [i]}(\eta_{m_i}),
\end{equation}
where $\eta_{m_i}$ is the matching conformal time between phases $i$ and $i+1$. This becomes $\eta_{i\uparrow}$ for sub-horizon and $\eta_{i\downarrow}$ for super-horizon.

Using these matching conditions, we can compute the sub-horizon matching matrix $M_{i\uparrow}$ and the super-horizon matching matrix $M_{i\downarrow}$ as follows:
\begin{enumerate}
    \item {\bf Sub-horizon matching matrix $M_{i\uparrow}$}~:

    If the modes with $k$ are matched at the boundary between phases $i$ and $i+1$ in the sub-horizon limit (for example, see the boundary between Phase II and III in Fig.~\ref{fig:Infboure.pdf}), we define the sub-horizon matching matrix $M_{i\uparrow}$ as 
    \begin{equation}
       \begin{pmatrix}
       A_{i+1}\\
       B_{i+1}
    \end{pmatrix} = M_{i\uparrow}
    \begin{pmatrix}
       A_i\\
       B_i
    \end{pmatrix}~,\quad k\eta_{i\uparrow}\gg 1~.
    \end{equation}

    In the sub-horizon limit ($k\eta_{i\uparrow}\gg 1$), using Eqs.~\ref{eqsi:matconone} and \ref{eqsi:matcontwo} with Eq.~\ref{eqsi:gensolsub}, we have  
    \begin{equation}
      K_{i+1\uparrow}\begin{pmatrix}
       A_{i+1} \\
       B_{i+1}
      \end{pmatrix} = K_{i\uparrow}\begin{pmatrix}
       A_i \\
       B_i
     \end{pmatrix},
   \end{equation}
   with a coefficient matrix $K_{i\uparrow}$ given by
   \begin{equation}
    K_{i\uparrow} \equiv \begin{pmatrix}
    e^{i \left( k \eta_{i\uparrow} - \frac{\tilde{\nu}_i \pi}{2} - \frac{\pi}{4} \right) } & e^{-i \left( k \eta_{i\uparrow} - \frac{\tilde{\nu}_i \pi}{2} - \frac{\pi}{4} \right) } \\
    \left(-\frac{\tilde{\nu}_i}{\eta_{i\uparrow}}+ik\right)e^{i \left( k \eta_{i\uparrow} - \frac{\tilde{\nu}_i \pi}{2} - \frac{\pi}{4} \right) } & \left(-\frac{\tilde{\nu}_i}{\eta_{i\uparrow}}-ik\right)e^{-i \left( k \eta_{i\uparrow} - \frac{\tilde{\nu}_i \pi}{2} - \frac{\pi}{4} \right) }
    \end{pmatrix}~.
    \end{equation}
    Using this, we obtain 
    \begin{equation}\label{eqsi:matchmatrup}
     M_{i\uparrow} = K_{i+1\uparrow}^{-1}K_{i\uparrow} = \begin{pmatrix}
         e^{-i(\tilde{\nu}_i - \tilde{\nu}_{i+1})\pi/2} & 0 \\
         0 & e^{i(\tilde{\nu}_i - \tilde{\nu}_{i+1})\pi/2}
     \end{pmatrix},
    \end{equation}
    where the approximation $\left(-\frac{\tilde{\nu}_i}{\eta_{i\uparrow}} \pm ik\right) \simeq \pm ik$ is used in the limit $k\eta_{i\uparrow}\gg 1$. $M_{i\uparrow}$ represents a phase shift and reduces to $\begin{pmatrix}
        1 & 0 \\
        0 & 1
    \end{pmatrix}$ when $\tilde{\nu}_{i+1} = \tilde{\nu}_i$.

    \item {\bf Super-horizon matching matrix $M_{i\downarrow}$}~:

    If the modes with $k$ are matched at the boundary between phases $i$ and $i+1$ in the super-horizon limit (for example, see the boundaries between Phase I and II or between Phase III and IV in Fig.~\ref{fig:Infboure.pdf}), we define the super-horizon matching matrix $M_{i\downarrow}$ as 
    \begin{equation}
       \begin{pmatrix}
        E_{i+1}\\
        F_{i+1}
       \end{pmatrix}=M_{i\downarrow}
       \begin{pmatrix}
        E_i\\
        F_i
       \end{pmatrix}~,\quad k\eta_{i\downarrow}\ll 1~.
    \end{equation}
    
    In the super-horizon limit ($k\eta_{i\downarrow}\ll 1$), using Eqs.~\ref{eqsi:matconone} and \ref{eqsi:matcontwo} with Eq.~\ref{eqsi:gensolsup}, we have 
    \begin{equation}
       K_{i+1\downarrow}\begin{pmatrix}
       E_{i+1}\\
       F_{i+1}
       \end{pmatrix}=K_{i\downarrow}
       \begin{pmatrix}
       E_i\\
       F_i
       \end{pmatrix}~,
    \end{equation}
    with a coefficient matrix $K_{i\downarrow}$ given by
    \begin{equation}
    K_{i\downarrow}=
    \begin{pmatrix}        
    \left(k\eta_{i\downarrow}\right)^{\tilde{\nu}_i}&\left(k\eta_{i\downarrow}\right)^{-\tilde{\nu}_i}\\
    \chi_i\left(k\eta_{i\downarrow}\right)^{\tilde{\nu}_i}& \bar{\chi}_i\left(k\eta_{i\downarrow}\right)^{-\tilde{\nu}_i}
    \end{pmatrix}~.
    \end{equation}

    Using this, we obtain
    \begin{align}\label{eqsi:matchmatrdown}
    M_{i\downarrow}=K_{i+1\downarrow}^{-1}K_{i\downarrow}
    =\frac{1}{\bar{\chi}_{i+1}-\chi_{i+1}}
    \begin{pmatrix}
        \left(\bar{\chi}_{i+1}-\chi_i\right)\left(k\eta_{i\downarrow}\right)^{\left(\tilde{\nu}_i-\tilde{\nu}_{i+1}\right)}&\left(\bar{\chi}_{i+1}-\bar{\chi}_i\right)\left(k\eta_{i\downarrow}\right)^{-\left(\tilde{\nu}_i+\tilde{\nu}_{i+1}\right)}\\
        \left(-\chi_{i+1}+\chi_i\right)\left(k\eta_{i\downarrow}\right)^{\left(\tilde{\nu}_i+\tilde{\nu}_{i+1}\right)}&\left(-\chi_{i+1}+\bar{\chi}_i\right)\left(k\eta_{i\downarrow}\right)^{-\left(\tilde{\nu}_i-\tilde{\nu}_{i+1}\right)}
    \end{pmatrix}~.
    \end{align}

\end{enumerate}

\subsection{Primordial Gravitational Wave Spectrum}

The spectrum of primordial gravitational waves after re-entering the horizon in the final phase of a generic big bounce cosmology (Phase IV) can be expressed as follows\footnote{
Using the unpolarized assumption, $\left|u_+\right|^2 = \left|u_\times\right|^2 = \left|u_k^{[4]}\right|^2$, we have 
\begin{equation}
    \left|h_r\right|^2 = \frac{16\pi G}{a^2}\left|u_r\right|^2 = \frac{16\pi G}{a^2}\left(\left|u_+\right|^2 + \left|u_\times\right|^2\right) = \frac{32\pi G}{a^2}\left|u_k^{[4]}\right|^2,
\end{equation}
with
\begin{align}
    \left|u_k^{[4]}\right|^2 &= \eta \left[\left|A_4\right|^2 \left|H_{\tilde{\nu}_4}^{(1)}\right|^2 + A_4^\ast B_4 H_{\tilde{\nu}_4}^{(1)\ast} H_{\tilde{\nu}_4}^{(2)} + B_4^\ast A_4 H_{\tilde{\nu}_4}^{\ast(2)} H_{\tilde{\nu}_4}^{(1)} + \left|B_4\right|^2 \left|H_{\tilde{\nu}_4}^{(2)}\right|^2\right] \\
    &\simeq \eta \left[\left|A_4\right|^2 + \left|B_4\right|^2\right] = \eta \begin{pmatrix}
        A_4^\ast, & B_4^\ast
    \end{pmatrix}
    \begin{pmatrix}
        A_4 \\
        B_4
    \end{pmatrix},
\end{align}
where we have ignored the oscillating terms that average out, $A_4^\ast B_4 H_{\tilde{\nu}_4}^{(1)\ast} H_{\tilde{\nu}_4}^{(2)} + B_4^\ast A_4 H_{\tilde{\nu}_4}^{\ast(2)} H_{\tilde{\nu}_4}^{(1)}$, and used Eq.~\ref{eqsi:gensolsub} for $k\eta > 1$ to simplify the expression in the last line. 
}

\begin{equation}\label{eqsi:a4b4}
    \mathcal{P}_{h} \equiv \frac{k^3|h_r|^2}{2\pi^2}
    \simeq \frac{k^3}{2\pi^2} \frac{32\pi G}{a^2} \eta \begin{pmatrix}
        A_4^\ast, & B_4^\ast
    \end{pmatrix}
    \begin{pmatrix}
        A_4 \\
        B_4
    \end{pmatrix},
\end{equation}
where $(A_4, B_4)$ are the coefficients of the solution in Phase IV, and the oscillating terms that average out in the spectrum have been ignored.

Using the transformation matrix $T_i$ and its inverse $T_i^{-1}$, and the matching matrices $M_{i\uparrow}$ and $M_{i\downarrow}$ derived in the last subsection, we can determine $(A_4, B_4)$ from $(A_1, B_1)$ as follows:
\begin{align}
    \begin{pmatrix}
    A_4\\
    B_4
\end{pmatrix}&=T_4^{-1}\begin{pmatrix}
    E_4\\
    F_4
\end{pmatrix}=T_4^{-1}M_{3\downarrow}\begin{pmatrix}
    E_3\\
    F_3
\end{pmatrix}=T_4^{-1}M_{3\downarrow}T_3\begin{pmatrix}
    A_3\\
    B_3
\end{pmatrix}\\
&=T_4^{-1}M_{3\downarrow}T_3M_{2\uparrow}\begin{pmatrix}
    A_2\\
    B_2
\end{pmatrix}=T_4^{-1}M_{3\downarrow}T_3M_{2\uparrow}T_2^{-1}\begin{pmatrix}
    E_2\\
    F_2
\end{pmatrix}\\
&=T_4^{-1}M_{3\downarrow}T_3M_{2\uparrow}T_2^{-1}M_{1\downarrow}\begin{pmatrix}
    E_1\\
    F_1
\end{pmatrix}=T_4^{-1}M_{3\downarrow}T_3M_{2\uparrow}T_2^{-1}M_{1\downarrow}T_1\begin{pmatrix}
    A_1\\
    B_1
\end{pmatrix}~,
\end{align}
where $(A_1, B_1)$ is determined by the initial quantum fluctuations of the Bunch-Davies vacuum in Phase I,
\begin{equation}\label{eqsi:a1b1exp}
\begin{pmatrix}
A_1\\
B_1
\end{pmatrix}=
\begin{pmatrix}
0\\
\frac{\sqrt{\pi}}{2} e^{-i\left(\frac{\Tilde{\nu}_1\pi}{2}+\frac{\pi}{4}\right)}
\end{pmatrix},
\end{equation}
the transformation matrices  for each phases take
\begin{equation}
    \quad T_1= 
    \begin{pmatrix}
    \alpha_1 & \alpha_1^\ast \\
    \beta_1 &\beta_1^\ast
    \end{pmatrix}~,\quad 
    T_3= 
    \begin{pmatrix}
    \alpha_3 & \alpha_3^\ast \\
    \beta_3 &\beta_3^\ast
    \end{pmatrix}~,
\end{equation}
and the inverse of transformation matrices take
\begin{equation}
    T_2^{-1}=\frac{1}{\alpha_2\beta_2^\ast-\alpha_2^\ast\beta_2}
    \begin{pmatrix}
        \beta_2^\ast &-\alpha_2^\ast\\
        -\beta_2&\alpha_2
    \end{pmatrix}~,\quad
    T_4^{-1}=\frac{1}{\alpha_4\beta_4^\ast-\alpha_4^\ast\beta_4}
    \begin{pmatrix}
        \beta_4^\ast &-\alpha_4^\ast\\
        -\beta_4&\alpha_4
    \end{pmatrix}~,
\end{equation}
the sub-horizon matching matrix takes
    \begin{equation}
     M_{2\uparrow}
     =\begin{pmatrix}
         e^{-i(\Tilde{\nu}_2-\Tilde{\nu}_3)\pi/2}&0\\
         0& e^{i(\Tilde{\nu}_2-\Tilde{\nu}_3)\pi/2}
     \end{pmatrix}~,
    \end{equation}
the super-horizon matching matrices, $M_{1\downarrow}$ and $M_{3\downarrow}$, take
    \begin{equation}
    M_{1\downarrow\le \frac{1}{2}}=
    \begin{pmatrix}
        \frac{\Tilde{\nu}_1}{\Tilde{\nu}_2}\left(k\eta_{1\downarrow}\right)^{\left(\Tilde{\nu}_1-\Tilde{\nu}_{2}\right)}&0\\
        \left(1-\frac{\Tilde{\nu}_1}{\Tilde{\nu}_2}\right)\left(k\eta_{1\downarrow}\right)^{\left(\Tilde{\nu}_1+\Tilde{\nu}_{2}\right)}&\left(k\eta_{1\downarrow}\right)^{-\left(\Tilde{\nu}_1-\Tilde{\nu}_{2}\right)}
    \end{pmatrix}~,\quad \nu_1\le \frac{1}{2}~,
    \end{equation}
        \begin{equation}
    M_{1\downarrow>\frac{1}{2}}=
    \begin{pmatrix}
        0&-\frac{\Tilde{\nu}_1}{\Tilde{\nu}_2}\left(k\eta_{1\downarrow}\right)^{-\left(\Tilde{\nu}_1+\Tilde{\nu}_{2}\right)}\\
        \left(k\eta_{1\downarrow}\right)^{\left(\Tilde{\nu}_1+\Tilde{\nu}_{2}\right)}&\left(1+\frac{\Tilde{\nu}_1}{\Tilde{\nu}_2}\right)\left(k\eta_{1\downarrow}\right)^{-\left(\Tilde{\nu}_1-\Tilde{\nu}_{2}\right)}
    \end{pmatrix}~,\quad \nu_1> \frac{1}{2}~.
    \end{equation}
    and 
    \begin{equation}
    M_{3\downarrow\le \frac{1}{2}}=
    \begin{pmatrix}
        \frac{\Tilde{\nu}_3}{\Tilde{\nu}_4}\left(k\eta_{3\downarrow}\right)^{\left(\Tilde{\nu}_3-\Tilde{\nu}_{4}\right)}&0\\
        \left(1-\frac{\Tilde{\nu}_3}{\Tilde{\nu}_4}\right)\left(k\eta_{3\downarrow}\right)^{\left(\Tilde{\nu}_3+\Tilde{\nu}_{4}\right)}&\left(k\eta_{3\downarrow}\right)^{-\left(\Tilde{\nu}_3-\Tilde{\nu}_{4}\right)}
    \end{pmatrix}~,\quad \nu_4\le \frac{1}{2},
   \end{equation}
   \begin{equation}
    M_{3\downarrow>\frac{1}{2}}=
    \begin{pmatrix}
        \left(1+\frac{\Tilde{\nu}_3}{\Tilde{\nu}_4}\right)\left(k\eta_{3\downarrow}\right)^{\left(\Tilde{\nu}_3-\Tilde{\nu}_{4}\right)}&\left(k\eta_{3\downarrow}\right)^{-\left(\Tilde{\nu}_3+\Tilde{\nu}_{4}\right)}\\
        -\frac{\Tilde{\nu}_3}{\Tilde{\nu}_4}\left(k\eta_{3\downarrow}\right)^{\left(\Tilde{\nu}_3+\Tilde{\nu}_{4}\right)}&0
    \end{pmatrix}~,\quad \nu_4>\frac{1}{2}.
\end{equation}

To derive the above results, we have used the algebra for $\chi_i$ and $\bar{\chi}_i$ based on Eqs.~\ref{eqsi:chiidef} and \ref{eqsi:bchiidef}, listed as follows:
\begin{enumerate}
    \item The values of $\chi_i$ and $\bar{\chi}_i$ for different $\nu_i$:
    \begin{align}\label{eq:chibcge}
    \chi_i &= 0, \quad \bar{\chi}_i = -2\tilde{\nu}_i, \quad \nu_i > \frac{1}{2}; \\
    \chi_i &=  2\tilde{\nu}_i, \quad \bar{\chi}_i = 0, \qquad \nu_i \le \frac{1}{2}. \label{eq:chibcle}
    \end{align}

    \item The orthogonality between $\chi_i$ and $\bar{\chi}_i$:
    \begin{equation}
    \chi_i \bar{\chi}_i = 0, \quad \mathrm{for~any}~\nu_i.
    \end{equation}

    \item The values of $\chi_i \chi_i$ and $\bar{\chi}_i \bar{\chi}_i$:
    \begin{align}
    \chi_i \chi_i = 0, \quad \bar{\chi}_i \bar{\chi}_i = 4\tilde{\nu}_i^2, &\quad \nu_i > \frac{1}{2}; \\
    \chi_i \chi_i = 4\tilde{\nu}_i^2, \quad \bar{\chi}_i \bar{\chi}_i = 0,  &\quad \nu_i \le \frac{1}{2}.
    \end{align}
\end{enumerate}

For simplifying representations of matrices, we define
\begin{equation}
X\equiv T_4^{-1}M_{3\downarrow}T_3M_{2\uparrow}T_2^{-1}M_{1\downarrow}T_1~,
\end{equation}
and 
\begin{equation}
     N\equiv X^\dagger X~,
\end{equation}
to get
\begin{equation}\label{eqsi:4x14n1}
    \begin{pmatrix}
        A_4\\
        B_4
    \end{pmatrix}=X
    \begin{pmatrix}
        A_1\\
        B_1
    \end{pmatrix}~,\quad 
    (A_4^\ast, B_4^\ast)
    \begin{pmatrix}
    A_4\\
    B_4
    \end{pmatrix}=
    (A_1^\ast, B_1^\ast)N
    \begin{pmatrix}
    A_1\\
    B_1
    \end{pmatrix}~.
\end{equation}

Using Eq.~\ref{eqsi:a1b1exp} and Eq.~\ref{eqsi:4x14n1}, the spectrum of primordial gravitational waves, Eq.~\ref{eqsi:a4b4}, can be simplified as
\begin{equation}\label{eqsi:phn22}
    \mathcal{P}_{h}\equiv \frac{k^3|h_r|^2}{2\pi^2}
    \simeq \frac{k^3}{2\pi^2}\frac{32\pi G}{a^2}\eta(A_1^\ast, B_1^\ast)X^\dagger X
    \begin{pmatrix}
    A_1\\
    B_1
    \end{pmatrix}=\frac{(k\eta)^3}{2\pi^2}\frac{\pi}{\nu_4^2}\frac{H^2}{m_p^2} N_{22}~,
\end{equation}
where $(aH)_i^{-1}=\nu_i^{-1}\eta$ and $m_p\equiv \sqrt{8\pi G}$ are used, and $N_{22}$ is the $_{22}$ component $N$.

\section{Four Categories of Big Bounce Cosmology and Model Demonstration}
Depending on the values of $\nu_1$ and $\nu_4$, $X$ can be categorized into four types as listed in Table~\ref{tab:xmatf}. Accordingly, $N$ can also be categorized into four types as listed in Table~\ref{tab:xmatf}.
\begin{table}[ht]
\centering
\begin{tabular}{|c|c|c|}
\hline
$X$ &\textbf{$\nu_1 \leq \frac{1}{2}$} & \textbf{$\nu_1 > \frac{1}{2}$} \\
\hline
$\nu_4 \leq \frac{1}{2}$& $X_{\le\le}\equiv T_4^{-1}M_{3\downarrow\le \frac{1}{2}}T_3M_{2\uparrow}T_2^{-1}M_{1\downarrow\le \frac{1}{2}}T_1$& $X_{>\le}\equiv T_4^{-1}M_{3\downarrow\le \frac{1}{2}}T_3M_{2\uparrow}T_2^{-1}M_{1\downarrow>\frac{1}{2}}T_1$  \\
\hline
$\nu_4 > \frac{1}{2}$& $X_{\le>}\equiv T_4^{-1}M_{3\downarrow> \frac{1}{2}}T_3M_{2\uparrow}T_2^{-1}M_{1\downarrow\le \frac{1}{2}}T_1$ & $X_{>>}\equiv T_4^{-1}M_{3\downarrow>\frac{1}{2}}T_3M_{2\uparrow}T_2^{-1}M_{1\downarrow>\frac{1}{2}}T_1$\\
\hline
\end{tabular}
\caption{Four categories of $X$ matrix}
\label{tab:xmatf}
\end{table}

\begin{table}[ht]
\centering
\begin{tabular}{|c|c|c|}
\hline
$N$ &\textbf{$\nu_1 \leq \frac{1}{2}$} & \textbf{$\nu_1 > \frac{1}{2}$} \\
\hline
$\nu_4 \leq \frac{1}{2}$& $N_{\le\le}\equiv X_{\le\le}^\dagger X_{\le\le}$ & $N_{>\le}\equiv X_{>\le}^\dagger X_{>\le}$  \\
\hline
$\nu_4 > \frac{1}{2}$& $N_{\le>}\equiv X_{\le>}^\dagger X_{\le>}$ & $N_{>>}\equiv X_{>>}^\dagger X_{>>}$\\
\hline
\end{tabular}
\caption{Four categories of $N\equiv X^\dagger X$ matrix}
\label{tab:nmatf}
\end{table}

\begin{figure}[htbp]
\centering 
\includegraphics[width=0.9\textwidth]{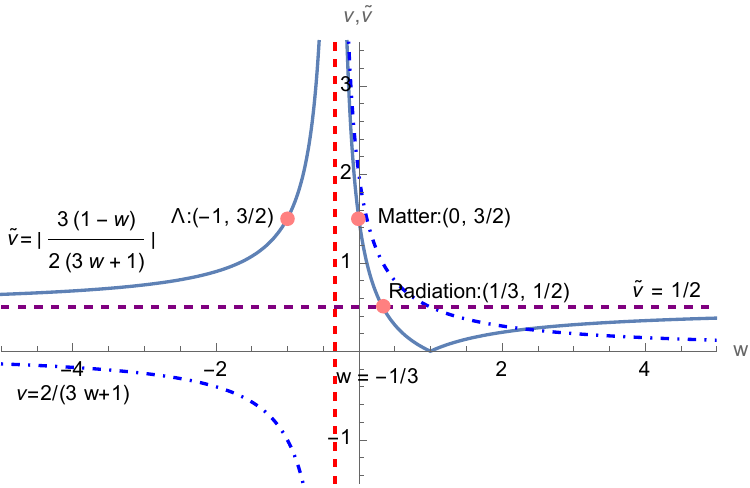}
\caption{\label{fig:wnutnu.pdf} Plot of $\nu = \frac{2}{3w+1}$ (dot-dashed blue line) and $\tilde{\nu} = \left| \frac{3(1-w)}{2(3w+1)} \right|$ (solid line). The horizontal dashed purple line represents $\tilde{\nu} = \frac{1}{2}$, and the vertical dashed red line indicates $w = -\frac{1}{3}$. Key points are marked on the $\tilde{\nu}$ curve: $\Lambda$ ($w = -1$, $\tilde{\nu} = \frac{3}{2}$), representing the cosmological constant-dominated era; Matter ($w = 0$, $\tilde{\nu} = \frac{3}{2}$), representing the matter-dominated era; and Radiation ($w = \frac{1}{3}$, $\tilde{\nu} = \frac{1}{2}$), representing the radiation-dominated era.}
\end{figure}

Now we introduce corresponding models to demonstrate each category. First, we relate each case to a physical cosmic background. Using the relation between the equation of state $w$ and the power law index $\nu$,
\begin{equation}
    \nu=\frac{2}{3w+1}~,
\end{equation}
and $\tilde{\nu}=\left|\nu-1/2\right|$, we have
\begin{equation}
    \Tilde{\nu}=\left|\frac{3(1-w)}{2(3w+1)}\right|~.
\end{equation}
In Fig.~\ref{fig:wnutnu.pdf}, we illustrate $\nu$ and $\tilde{\nu}$ in terms of the EoS of the cosmic background $w$, which can help us to quickly figure out each case.

Next, we present the four toy models listed in Table~\ref{tab:modelmatf} for each category, demonstrating that our analytical results of big bounce cosmology can be straightforwardly applied to various bouncing universe models.

\begin{table}[ht]
\centering
\begin{tabular}{|c|c|c|}
\hline
Model &\textbf{$\nu_1 \leq \frac{1}{2}$} & \textbf{$\nu_1 > \frac{1}{2}$} \\
\hline
$\nu_4 \leq \frac{1}{2}$& Model-3 & Model-4  \\
\hline
$\nu_4 > \frac{1}{2}$& Model-1& Model-2\\
\hline
\end{tabular}
\caption{Four models for each category}
\label{tab:modelmatf}
\end{table}

\begin{enumerate}
    \item Model-1: $(w_1, w_2, w_3, w_4) = \left(\infty, -\infty, -\infty, \frac{1}{3}\right)$.

    In Model-1, we assume that in Phase I, the universe initially undergoes a collapsing contraction ($\dot{a}<0, \ddot{a}<0$) dominated by some exotic matter ($w \rightarrow +\infty$), which corresponds to $\nu_1 \rightarrow 0 \le \frac{1}{2}$ and $\tilde{\nu}_1 = \frac{1}{2}$. In Phases II and III, the universe undergoes bouncing contraction ($\dot{a}<0, \ddot{a}>0$) and bouncing expansion ($\dot{a}>0, \ddot{a}>0$) dominated by bouncing fields ($w_2 = w_3 \rightarrow -\infty$), which corresponds to $\nu_2 = \nu_3 = 0$ and $\tilde{\nu}_2 = \tilde{\nu}_3 = \frac{1}{2}$. In Phase IV, the universe undergoes decelerating expansion ($\dot{a}>0, \ddot{a}<0$) dominated by standard post-reheating radiation ($w_4 = \frac{1}{3}$), which corresponds to $\nu_4 = 1 > \frac{1}{2}$ and $\tilde{\nu}_4 = \frac{1}{2}$.

    Thus, for Model-1, we have 
    \begin{equation}\label{eqsi:m1nutnu}
        (\nu_1,\nu_2,\nu_3,\nu_4)=(0,0,0,1)~, \quad (\Tilde{\nu}_1,\Tilde{\nu}_2,\Tilde{\nu}_3,\Tilde{\nu}_4)=\left(\frac{1}{2},\frac{1}{2},\frac{1}{2},\frac{1}{2}\right)~,
    \end{equation} 
    which corresponds to $N_{\le>}$ in Tab.~\ref{tab:nmatf}. Substituting Eq.~\ref{eqsi:m1nutnu} into $X_{\le>}$ and $N_{\le >}$ in Tab.~\ref{tab:xmatf} and Tab.~\ref{tab:nmatf}, we obtain
    \begin{equation}
        X_{\le >\mathrm{Model-1}}=
        \begin{pmatrix}
        1 - \frac{i}{2k\eta_{3\downarrow}} - \frac{ik\eta_{3\downarrow}}{2} & -\frac{i (i + k\eta_{3\downarrow})^2}{2k\eta_{3\downarrow}} \\
        \frac{i (-i + k\eta_{3\downarrow})^2}{2k\eta_{3\downarrow}} & 1 + \frac{i}{2k\eta_{3\downarrow}} + \frac{ik\eta_{3\downarrow}}{2}
        \end{pmatrix}
    \end{equation}
    and
    \begin{align}
        N_{\le >\mathrm{Model-1}}&=X^\dagger_{\le >\mathrm{Model-1}}X_{\le >\mathrm{Model-1}}\\
        &=
        \begin{pmatrix}
        \frac{1}{2} \left(4 + \frac{1}{(k\eta_{3\downarrow})^2} + (k\eta_{3\downarrow})^2\right) & \frac{-1 + 4i k\eta_{3\downarrow} + 4 (k\eta_{3\downarrow})^2 + (k\eta_{3\downarrow})^4}{2 (k\eta_{3\downarrow})^2} \\
        \frac{-1 - 4i k\eta_{3\downarrow} + 4 (k\eta_{3\downarrow})^2 + (k\eta_{3\downarrow})^4}{2 (k\eta_{3\downarrow})^2} & \frac{1}{2} \left(4 + \frac{1}{(k\eta_{3\downarrow})^2} + (k\eta_{3\downarrow})^2\right)
        \end{pmatrix}\\
        &\simeq 
        \begin{pmatrix}
         \frac{1}{2 (k\eta_{3\downarrow})^2}  & -\frac{1}{2 (k\eta_{3\downarrow})^2}   \\
         -\frac{1}{2 (k\eta_{3\downarrow})^2}  & \frac{1}{2 (k\eta_{3\downarrow})^2}
        \end{pmatrix}
        =\frac{1}{2 (k\eta_{3\downarrow})^2} 
        \begin{pmatrix}
            1&-1\\
            -1&1
        \end{pmatrix}~,\quad k\eta_{3\downarrow}\ll 1~.
    \end{align}
    Using Eq.~\ref{eqsi:phn22}, we obtain
    \begin{equation}
        \mathcal{P}_{h~\mathrm{Model-1}}\simeq \frac{1}{2\pi}\frac{H^2_\ast}{m_p^2}\cdot \frac{1}{2 (k\eta_{3\downarrow})^2} ,\quad k \eta_\ast\simeq 1~.
    \end{equation}

    \item Model-2: $(w_1, w_2, w_3, w_4) = \left(0, -\infty, -\infty, \frac{1}{3}\right)$.

    In Model-2, we assume that in Phase I, the universe initially undergoes a collapsing contraction ($\dot{a}<0, \ddot{a}<0$) dominated by cold matter ($w = 0$), akin to the matter bounce universe model, which corresponds to $\nu_1 = 2 > \frac{1}{2}$ and $\tilde{\nu}_1 = \frac{3}{2}$. In Phases II and III, the universe undergoes bouncing contraction ($\dot{a}<0, \ddot{a}>0$) and bouncing expansion ($\dot{a}>0, \ddot{a}>0$) dominated by bouncing fields ($w_2 = w_3 \rightarrow -\infty$), which corresponds to $\nu_2 = \nu_3 = 0$ and $\tilde{\nu}_2 = \tilde{\nu}_3 = \frac{1}{2}$. In Phase IV, the universe undergoes decelerating expansion ($\dot{a}>0, \ddot{a}<0$) dominated by standard post-reheating radiation ($w_4 = \frac{1}{3}$), which corresponds to $\nu_4 = 1 > \frac{1}{2}$ and $\tilde{\nu}_4 = \frac{1}{2}$.
    Thus, for Model-2, we have 
    \begin{equation}\label{eqsi:m2nutnu}
        (\nu_1,\nu_2,\nu_3,\nu_4)=(2,0,0,1)~,\quad (\Tilde{\nu}_1,\Tilde{\nu}_2,\Tilde{\nu}_3,\Tilde{\nu}_4)=\left(\frac{3}{2},\frac{1}{2},\frac{1}{2},\frac{1}{2}\right)~,
    \end{equation} 
    which corresponds to $N_{>>}$ in Tab.~\ref{tab:nmatf}. Substituting Eq.~\ref{eqsi:m2nutnu} into $X_{>>}$ and $N_{>>}$ in Tab.~\ref{tab:xmatf} and Tab.~\ref{tab:nmatf}, we obtain
    \begin{equation}
        X_{>>\mathrm{Model-2}}=
        \begin{pmatrix}
        \frac{-12ik\eta_{1\downarrow} + (k\eta_{1\downarrow})^4 + 9k\eta_{3\downarrow} (2i + k\eta_{3\downarrow})}{6 (k\eta_{1\downarrow})^2 k\eta_{3\downarrow}} & \frac{12ik\eta_{1\downarrow} + (k\eta_{1\downarrow})^4 - 9k\eta_{3\downarrow} (2i + k\eta_{3\downarrow})}{6 (k\eta_{1\downarrow})^2 k\eta_{3\downarrow}} \\
        \frac{-12ik\eta_{1\downarrow} + (k\eta_{1\downarrow})^4 - 9k\eta_{3\downarrow} (-2i + k\eta_{3\downarrow})}{6 (k\eta_{1\downarrow})^2 k\eta_{3\downarrow}} & \frac{12ik\eta_{1\downarrow} + (k\eta_{1\downarrow})^4 + 9k\eta_{3\downarrow} (-2i + k\eta_{3\downarrow})}{6 (k\eta_{1\downarrow})^2 k\eta_{3\downarrow}}
        \end{pmatrix}~,
    \end{equation}
    and
    \begin{align}
        N_{>>\mathrm{Model-2}}=X^\dagger_{>>\mathrm{Model-2}}X_{>>\mathrm{Model-2}}=
        \begin{pmatrix}
            N_{11}&N_{12}\\
            N_{21}&N_{22}
        \end{pmatrix}~,
    \end{align}
    with 
    \begin{align}
        &N_{11}=
        \frac{144 (k\eta_{1\downarrow})^2 + (k\eta_{1\downarrow})^8 - 432 k\eta_{1\downarrow} k\eta_{3\downarrow} + 81 (k\eta_{3\downarrow})^2 (4 + (k\eta_{3\downarrow})^2)}{18 (k\eta_{1\downarrow})^4 (k\eta_{3\downarrow})^2}~,
        \\
        &N_{12}=\frac{-144 (k\eta_{1\downarrow})^2 + 24 i (k\eta_{1\downarrow})^5 + (k\eta_{1\downarrow})^8 + 432 k\eta_{1\downarrow} k\eta_{3\downarrow} - 36 i (k\eta_{1\downarrow})^4 k\eta_{3\downarrow} - 81 (k\eta_{3\downarrow})^2 (4 + (k\eta_{3\downarrow})^2)}{18 (k\eta_{1\downarrow})^4 (k\eta_{3\downarrow})^2}~, 
        \\
        &N_{21}=
        \frac{-144 (k\eta_{1\downarrow})^2 - 24 i (k\eta_{1\downarrow})^5 + (k\eta_{1\downarrow})^8 + 432 k\eta_{1\downarrow} k\eta_{3\downarrow} + 36 i (k\eta_{1\downarrow})^4 k\eta_{3\downarrow} - 81 (k\eta_{3\downarrow})^2 (4 + (k\eta_{3\downarrow})^2)}{18 (k\eta_{1\downarrow})^4 (k\eta_{3\downarrow})^2}~,
        \\
        &N_{22}=\frac{144 (k\eta_{1\downarrow})^2 + (k\eta_{1\downarrow})^8 - 432 k\eta_{1\downarrow} k\eta_{3\downarrow} + 81 (k\eta_{3\downarrow})^2 (4 + (k\eta_{3\downarrow})^2)}{18 (k\eta_{1\downarrow})^4 (k\eta_{3\downarrow})^2}~.\label{eqsi:n22m2}
    \end{align}
    Using Eq.~\ref{eqsi:phn22} and Eq.~\ref{eqsi:n22m2}, we obtain
    \begin{equation}
        \mathcal{P}_{h~\mathrm{Model-2}}\simeq \frac{1}{2\pi}\frac{H^2_\ast}{m_p^2}\cdot \frac{144 (k\eta_{1\downarrow})^2 - 432 k\eta_{1\downarrow} k\eta_{3\downarrow} + 324 (k\eta_{3\downarrow})^2 }{18 (k\eta_{1\downarrow})^4 (k\eta_{3\downarrow})^2} ,\quad k \eta_\ast\simeq 1~,
    \end{equation}
    where we note that, in general, the spectrum is not only dependent on $\eta_{3\downarrow}$ but also on $\eta_{1\downarrow}$.

    For further simplification, we can assume that the bouncing process is symmetric ($\eta_{\downarrow s} \equiv \eta_{1\downarrow} = \eta_{3\downarrow}$). The result can then be simplified as
    \begin{align}
        N_{>>\mathrm{Model-2}s}=
        \begin{pmatrix}
        \frac{2}{(k\eta_{\downarrow s})^4} & -\frac{2}{(k\eta_{\downarrow s})^4}  \\
        -\frac{2}{(k\eta_{\downarrow s})^4} & \frac{2}{(k\eta_{\downarrow s})^4} 
        \end{pmatrix}
        =\frac{2}{ (k\eta_{\downarrow s})^4} 
        \begin{pmatrix}
            1&-1\\
            -1&1
        \end{pmatrix}~,
    \end{align}
    which leads to 
    \begin{equation}
        \mathcal{P}_{h~\mathrm{Model-2} s}\simeq \frac{1}{2\pi}\frac{H^2_\ast}{m_p^2}\cdot \frac{2}{ (k\eta_{\downarrow s})^4} ,\quad k \eta_\ast\simeq 1~,
    \end{equation}
    where the subscript $_s$ labels the symmetric simplification.

    To illustrate the other two cases, $N_{\le \le}$ and $N_{>\le}$, we assume Phase IV is not the standard post-reheating radiation-dominated era. (Of course, eventually, Phase IV transitions to the standard post-reheating radiation-dominated era to accommodate astrophysical observations.) For simplicity, we take $w_4 = \frac{3}{2}$ for the following two models.

    \item Model-3: $(w_1, w_2, w_3, w_4) = \left(\infty, -\infty, -\infty, \frac{3}{2}\right)$.

    In Model-3, we assume that in Phase I, the universe initially undergoes a collapsing contraction ($\dot{a}<0, \ddot{a}<0$) dominated by some exotic matter ($w \rightarrow +\infty$), which corresponds to $\nu_1 \rightarrow 0 \le \frac{1}{2}$ and $\tilde{\nu}_1 = \frac{1}{2}$ (akin to Model-1). In Phases II and III, the universe undergoes bouncing contraction ($\dot{a}<0, \ddot{a}>0$) and bouncing expansion ($\dot{a}>0, \ddot{a}>0$) dominated by bouncing fields ($w_2 = w_3 \rightarrow -\infty$), which corresponds to $\nu_2 = \nu_3 = 0$ and $\tilde{\nu}_2 = \tilde{\nu}_3 = \frac{1}{2}$ (akin to Model-1). The difference is in Phase IV, where the universe undergoes decelerating expansion ($\dot{a}>0, \ddot{a}<0$) dominated by some exotic matter ($w_4 = \frac{3}{2}$), which corresponds to $\nu_4 = \frac{4}{11} < \frac{1}{2}$ and $\tilde{\nu}_4 = \frac{3}{22}$.
    Thus, for Model-3, we have 
    \begin{equation}\label{eqsi:m3nutnu}
        (\nu_1,\nu_2,\nu_3,\nu_4)=(0,0,0,\frac{4}{11})~, \quad (\Tilde{\nu}_1,\Tilde{\nu}_2,\Tilde{\nu}_3,\Tilde{\nu}_4)=\left(\frac{1}{2},\frac{1}{2},\frac{1}{2},\frac{3}{22}\right)~,
    \end{equation} 
    which corresponds to $N_{\le\le}$ in Tab.~\ref{tab:nmatf}. Substituting Eq.~\ref{eqsi:m3nutnu} into $X_{\le\le}$ and $N_{\le \le}$ in Tab.~\ref{tab:xmatf} and Tab.~\ref{tab:nmatf}, we obtain (the leading order for  $k\eta_{3\downarrow}\ll 1$)
    \begin{equation}
        X_{\le \le \mathrm{Model-3}}=
        \begin{pmatrix}
        -\frac{2^{4/11} \sqrt{\pi}}{(k\eta_{3\downarrow})^{4/11} \Gamma\left(\frac{3}{22}\right) \left(-1 + i \cos\left(\frac{5\pi}{22}\right) + \sin\left(\frac{5\pi}{22}\right)\right)} & \frac{2^{4/11} \sqrt{\pi}}{(k\eta_{3\downarrow})^{4/11} \Gamma\left(\frac{3}{22}\right) \left(-1 + i \cos\left(\frac{5\pi}{22}\right) + \sin\left(\frac{5\pi}{22}\right)\right)} \\
        \frac{2^{4/11} \sqrt{\pi} \left(-i \cos\left(\frac{5\pi}{22}\right) - \sin\left(\frac{5\pi}{22}\right)\right)}{(k\eta_{3\downarrow})^{4/11} \Gamma\left(\frac{3}{22}\right) \left(-1 + i \cos\left(\frac{5\pi}{22}\right) + \sin\left(\frac{5\pi}{22}\right)\right)} & \frac{2^{4/11} \sqrt{\pi} \left(i \cos\left(\frac{5\pi}{22}\right) + \sin\left(\frac{5\pi}{22}\right)\right)}{(k\eta_{3\downarrow})^{4/11} \Gamma\left(\frac{3}{22}\right) \left(-1 + i \cos\left(\frac{5\pi}{22}\right) + \sin\left(\frac{5\pi}{22}\right)\right)}
\end{pmatrix}
    \end{equation}
    and
    \begin{align}
        &N_{\le \le \mathrm{Model-3}}=X^\dagger_{\le \le\mathrm{Model-3}}X_{\le \le\mathrm{Model-3}}\\
        &=
       \begin{pmatrix}
       \frac{2^{8/11} \pi}{(k\eta_{3\downarrow})^{8/11} \Gamma\left(\frac{3}{22}\right)^2 \left(\cos\left(\frac{5\pi}{44}\right) - \sin\left(\frac{5\pi}{44}\right)\right)^2} & -\frac{2^{8/11} \pi}{(k\eta_{3\downarrow})^{8/11} \Gamma\left(\frac{3}{22}\right)^2 \left(\cos\left(\frac{5\pi}{44}\right) - \sin\left(\frac{5\pi}{44}\right)\right)^2} \\
      -\frac{2^{8/11} \pi}{(k\eta_{3\downarrow})^{8/11} \Gamma\left(\frac{3}{22}\right)^2 \left(\cos\left(\frac{5\pi}{44}\right) - \sin\left(\frac{5\pi}{44}\right)\right)^2} & \frac{2^{8/11} \pi}{(k\eta_{3\downarrow})^{8/11} \Gamma\left(\frac{3}{22}\right)^2 \left(\cos\left(\frac{5\pi}{44}\right) - \sin\left(\frac{5\pi}{44}\right)\right)^2}
      \end{pmatrix}
        \\
        &
        =\frac{2^{8/11} \pi}{(k\eta_{3\downarrow})^{8/11} \Gamma\left(\frac{3}{22}\right)^2 \left(\cos\left(\frac{5\pi}{44}\right) - \sin\left(\frac{5\pi}{44}\right)\right)^2}
        \begin{pmatrix}
            1&-1\\
            -1&1
        \end{pmatrix}\simeq \frac{\mathcal{C}_3}{(k\eta_{3\downarrow})^{8/11}}
        \begin{pmatrix}
            1&-1\\
            -1&1
        \end{pmatrix}~,
    \end{align}
    where $\mathcal{C}_3\equiv \frac{2^{8/11} \pi}{\Gamma\left(\frac{3}{22}\right)^2 \left(\cos\left(\frac{5\pi}{44}\right) - \sin\left(\frac{5\pi}{44}\right)\right)^2}=0.32$. 
    Using Eq.~\ref{eqsi:phn22}, we obtain
    \begin{equation}
        \mathcal{P}_{h~\mathrm{Model-3}}\simeq \frac{1}{2\pi}\frac{H^2_\ast}{m_p^2}\cdot \frac{\mathcal{C}_3}{(k\eta_{3\downarrow})^{8/11}}\cdot \left(\frac{11}{4}\right)^2 ,\quad k \eta_\ast\simeq 1~.
    \end{equation}

    \item Model-4: $(w_1, w_2, w_3, w_4) = \left(0, -\infty, -\infty, \frac{3}{2}\right)$.

    In Model-4, we assume that in Phase I, the universe initially undergoes a collapsing contraction ($\dot{a}<0, \ddot{a}<0$) dominated by cold matter ($w = 0$), which corresponds to $\nu_1 = 2 \ge \frac{1}{2}$ and $\tilde{\nu}_1 = \frac{3}{2}$. In Phases II and III, the universe undergoes bouncing contraction ($\dot{a}<0, \ddot{a}>0$) and bouncing expansion ($\dot{a}>0, \ddot{a}>0$) dominated by bouncing fields ($w_2 = w_3 \rightarrow -\infty$), which corresponds to $\nu_2 = \nu_3 = 0$ and $\tilde{\nu}_2 = \tilde{\nu}_3 = \frac{1}{2}$. In Phase IV, the universe undergoes decelerating expansion ($\dot{a}>0, \ddot{a}<0$) dominated by some exotic matter ($w_4 = \frac{3}{2}$), which corresponds to $\nu_4 = \frac{4}{11} < \frac{1}{2}$ and $\tilde{\nu}_4 = \frac{3}{22}$.
    Thus, for Model-4, we have 
    \begin{equation}\label{eqsi:m4nutnu}
        (\nu_1,\nu_2,\nu_3,\nu_4)=(2,0,0,\frac{4}{11})~, \quad (\Tilde{\nu}_1,\Tilde{\nu}_2,\Tilde{\nu}_3,\Tilde{\nu}_4)=\left(\frac{3}{2},\frac{1}{2},\frac{1}{2},\frac{3}{22}\right)~,
    \end{equation} 
    which corresponds to $N_{>\le}$ in Tab.~\ref{tab:nmatf}. Substituting Eq.~\ref{eqsi:m4nutnu} into $X_{>\le}$ and $N_{>\le}$ in Tab.~\ref{tab:xmatf} and Tab.~\ref{tab:nmatf}, we obtain (the leading order for the symmetric bounce $k\eta_{\downarrow s}\equiv k\eta_{1\downarrow}=k\eta_{3\downarrow}\ll 1$ for simplicity)
    \begin{equation}
        X_{>\le \mathrm{Model-4} s}=
        \begin{pmatrix}
        -\frac{11 \cdot 2^{7/11} \Gamma\left(\frac{25}{22}\right) \left(\cos\left(\frac{3\pi}{22}\right) + i \sin\left(\frac{3\pi}{22}\right)\right) \sin\left(\frac{3\pi}{22}\right)}{(k\eta_{\downarrow s})^{18/11} \sqrt{\pi} \left(-1 + i \cos\left(\frac{5\pi}{22}\right) + \sin\left(\frac{5\pi}{22}\right)\right)} & \frac{11 \cdot 2^{7/11} \Gamma\left(\frac{25}{22}\right) \left(\cos\left(\frac{3\pi}{22}\right) + i \sin\left(\frac{3\pi}{22}\right)\right) \sin\left(\frac{3\pi}{22}\right)}{(k\eta_{\downarrow s})^{18/11} \sqrt{\pi} \left(-1 + i \cos\left(\frac{5\pi}{22}\right) + \sin\left(\frac{5\pi}{22}\right)\right)} \\
        -\frac{11 \cdot 2^{7/11} \Gamma\left(\frac{25}{22}\right) \left(\cos\left(\frac{3\pi}{22}\right) + i \sin\left(\frac{3\pi}{22}\right)\right) \sin\left(\frac{3\pi}{22}\right)}{(k\eta_{\downarrow s})^{18/11} \sqrt{\pi} \left(-1 + i \cos\left(\frac{5\pi}{22}\right) + \sin\left(\frac{5\pi}{22}\right)\right)} & \frac{11 \cdot 2^{7/11} \Gamma\left(\frac{25}{22}\right) \left(\cos\left(\frac{3\pi}{22}\right) + i \sin\left(\frac{3\pi}{22}\right)\right) \sin\left(\frac{3\pi}{22}\right)}{(k\eta_{\downarrow s})^{18/11} \sqrt{\pi} \left(-1 + i \cos\left(\frac{5\pi}{22}\right) + \sin\left(\frac{5\pi}{22}\right)\right)}
\end{pmatrix}
    \end{equation}
    and
    \begin{align}
        &N_{>\le \mathrm{Model-4} s}=X^\dagger_{>\le \mathrm{Model-4} s}X_{>\le \mathrm{Model-4} s}\\
        &=\begin{pmatrix}
        \frac{242 \cdot 2^{3/11} \Gamma\left(\frac{25}{22}\right)^2 \sin^2\left(\frac{3\pi}{22}\right)}{(k\eta_{\downarrow s})^{36/11} \pi \left(\cos\left(\frac{5\pi}{44}\right) - \sin\left(\frac{5\pi}{44}\right)\right)^2} & -\frac{242 \cdot 2^{3/11} \Gamma\left(\frac{25}{22}\right)^2 \sin^2\left(\frac{3\pi}{22}\right)}{(k\eta_{\downarrow s})^{36/11} \pi \left(\cos\left(\frac{5\pi}{44}\right) - \sin\left(\frac{5\pi}{44}\right)\right)^2} \\
        -\frac{242 \cdot 2^{3/11} \Gamma\left(\frac{25}{22}\right)^2 \sin^2\left(\frac{3\pi}{22}\right)}{(k\eta_{\downarrow s})^{36/11} \pi \left(\cos\left(\frac{5\pi}{44}\right) - \sin\left(\frac{5\pi}{44}\right)\right)^2} & \frac{242 \cdot 2^{3/11} \Gamma\left(\frac{25}{22}\right)^2 \sin^2\left(\frac{3\pi}{22}\right)}{(k\eta_{\downarrow s})^{36/11} \pi \left(\cos\left(\frac{5\pi}{44}\right) - \sin\left(\frac{5\pi}{44}\right)\right)^2}
\end{pmatrix}\\
        &\simeq 
        \frac{242 \cdot 2^{3/11} \Gamma\left(\frac{25}{22}\right)^2 \sin^2\left(\frac{3\pi}{22}\right)}{(k\eta_{\downarrow s})^{36/11} \pi \left(\cos\left(\frac{5\pi}{44}\right) - \sin\left(\frac{5\pi}{44}\right)\right)^2} 
        \begin{pmatrix}
            1&-1\\
            -1&1
        \end{pmatrix}
        =\frac{\mathcal{C}_4}{(k\eta_{\downarrow s})^{36/11}}\begin{pmatrix}
            1&-1\\
            -1&1
        \end{pmatrix}~,
    \end{align}
    where $\mathcal{C}_{4}\equiv \frac{242 \cdot 2^{3/11} \Gamma\left(\frac{25}{22}\right)^2 \sin^2\left(\frac{3\pi}{22}\right)}{\pi \left(\cos\left(\frac{5\pi}{44}\right) - \sin\left(\frac{5\pi}{44}\right)\right)^2} =40.9$.
    Using Eq.~\ref{eqsi:phn22}, we obtain
    \begin{equation}
        \mathcal{P}_{h~\mathrm{Model-4}}\simeq \frac{1}{2\pi}\frac{H^2_\ast}{m_p^2}\cdot \frac{\mathcal{C}_4}{(k\eta_{\downarrow s})^{36/11}}\cdot \left(\frac{11}{4}\right)^2 ,\quad k \eta_\ast\simeq 1~.
    \end{equation}

\end{enumerate}

Clearly, these four toy models for each category of big bounce cosmology demonstrate that our analytical results can be straightforwardly applied to various bouncing universe models.

\section{Conclusion}
In this work, we present the explicit form of the primordial gravitational wave spectrum generated in a generic big bounce cosmology, paving the way to explore the initial stage of a generic bouncing universe with SGWB signals recently detected and searched by PTAs and upcoming advanced gravitational wave detectors. Based on our analytical results, we find that, according to the evolution of primordial gravitational waves, a generic scenario of big bounce cosmology can be categorized into four types. We introduce four toy models for these categories, demonstrating that our analytical results apply to various bouncing universe models with constant $w_i$.

For future applications of our results to interpret SGWB signals searched by PTAs and upcoming advanced gravitational wave detectors such as SKA, Taiji, Tianqin, LISA, DECIGO, and aLIGO/Virgo/KAGRA~\cite{Nan:2011um, Ruan:2018tsw, TianQin:2015yph, LIGOScientific:2014pky, VIRGO:2014yos, KAGRA:2018plz, LISA:2017pwj, Kawamura:2011zz, Punturo:2010zz}, one can straightforwardly obtain the primordial gravitational wave spectrum $\mathcal{P}_h$ from Eq.~\ref{eqsi:phn22} (and determine the spectral index $n_T$ from $\mathcal{P}_h$) for a given model with specified values of $(\nu_1,\nu_1,\nu_1,\nu_1,\eta_{1\downarrow},\eta_{3\downarrow})$. Using the general expression of the SGWB spectrum induced by primordial gravitational waves \cite{Caprini:2018mtu}, one can eventually obtain the SGWB spectrum for a given big bounce cosmology model, 
\begin{equation} \label{eqsi:sgwbph}
    \Omega_\mathrm{GW}(f)h^2=\frac{3}{128}\Omega_{\gamma 0}h^2\cdot \mathcal{P}_h(f)\left[\frac{1}{2}\left(\frac{f_\mathrm{eq}}{f}\right)^2+\frac{16}{9}\right]~, 
\end{equation}
which can be used to perform Bayesian analysis with datasets from PTAs and upcoming gravitational wave detectors, where $f = k/(2\pi a_0)$ is the frequency of SGWB signals today ($a_0 = 1$), $f_\mathrm{eq} = 2.01\times 10^{-17}~\mathrm{Hz}$ is the frequency today corresponding to matter-radiation equality, $\Omega_{\gamma 0} = 2.474 \times 10^{-5} h^{-2}$ is the energy density fraction of radiation today, and $h = 0.677$ is the reduced Hubble constant.

For future development of our results, exotic cosmic phases can also be taken into account in the regions between Phase I and Phase II and/or between Phase III and Phase IV, similar to how an additional phase dominated by exotic components is added between the inflation and post-reheating eras in conventional models. Such modifications could result in an enhancement or suppression of SGWB signals and diversify possible theoretical predictions of SGWB signals in big bounce cosmology. Of course, recent and upcoming gravitational wave searches can precisely constrain and test or falsify these theoretical predictions from big bounce cosmology.

\acknowledgments
C.L. is supported by the NSFC under Grants No.11963005 and No. 11603018, by Yunnan Provincial Foundation under Grants No.202401AT070459, No.2019FY003005, and No. 2016FD006, by Young and Middle-aged Academic and Technical Leaders in Yunnan Province Program, by Yunnan Provincial High level Talent Training Support Plan Youth Top Program, by Yunnan University Donglu Talent Young Scholar, and by the NSFC under Grant No.11847301 and by the Fundamental Research Funds for the Central Universities under Grant No. 2019CDJDWL0005.

 \bibliographystyle{JHEP}
 \bibliography{biblio.bib}



\end{document}